\documentclass[aps,preprint,floatfix,superscriptaddress,amssymb]{revtex4-1}

\usepackage[english]{babel}
\usepackage[utf8x]{inputenc}
\usepackage{amsmath}
\usepackage{array}
\usepackage{graphicx}% Include figure files
\usepackage{dcolumn}% Align table columns on decimal point
\usepackage{bm}% bold math
\usepackage[caption=false]{subfig}
\usepackage{todonotes}

\begin{document}
\title{Rational design principles for giant spin Hall effect in \textit{5d}-transition metal oxides}
\author{Priyamvada Jadaun}
\email{priyamvada@utexas.edu}
\author{Leonard F Register}
\author{Sanjay K Banerjee}
\affiliation{Department of Electrical and Computer Engineering, The University of Texas at Austin, Austin, Texas. USA.}
\date{\today}

\begin{abstract}
Spin Hall effect (SHE), a mechanism by which materials convert a \textit{charge} current into a \textit{spin} current, invokes interesting physics and promises to empower transformative, energy-efficient memory technology. However, fundamental questions remain about the essential factors that determine SHE. Here we solve this open problem, presenting a comprehensive theory of five \textit{foundational factors} that control the value of intrinsic SHE in transition metal oxides. Arising from our key insight regarding the inherently geometric nature of SHE, we demonstrate that two of these factors are crystal field strength and structural distortions. Moreover, we discover that a new class of materials (anti-perovskites) promises to demonstrate \textit{giant} SHE, that is an order of magnitude larger than that reported for any oxide. We derive three other factors that control SHE and demonstrate the nuanced role of electron correlations. Our findings bring deeper insight into the physics driving SHE, and could help enhance, as well as, externally control SHE values.
\end{abstract}

\maketitle

\section{Introduction}
A spintronic memory device that demonstrates all-electrical control of the magnetic state, is highly attractive for next-generation memory technology \cite{Yan_Large}. These  devices are a promising solution to our society's rapidly growing demand for data storage \cite{Sander_Roadmap}, owing to their non-volatility, quick random access, low degradation, energy-efficiency and ability to be integrated into conventional electrical circuits \cite{ Chappert_emergence, Wolf_Spintronics, Prinz_Magnetoelectronics}. A particularly promising design for these spintronic devices utilizes the spin Hall effect (SHE) \cite{Dyakonov_Current, Hirsch_Spin, Sinova_Spin_Hall, Valenzuela_Direct, Saitoh_Conversion, Kimura_Room}. Spin Hall effect (SHE) is a mechanism by which a material converts an incoming charge current into a spin current. The resultant spin current is subsequently used to control and switch the magnetic state of a ferromagnet, via spin-orbit torques, ultimately leading to the highly desired, electrical control of magnetism. As a result, SHE promises significant technological impact.

Along with its strong technological importance, SHE is also of fundamental interest owing to its rich physics. However, important questions still remain about the essential factors that determine SHE. Here we address this open problem, presenting a comprehensive theory of the fundamental factors that control the value of intrinsic SHE in transition metal oxides (TMOs). Our findings stem, in part, from the key realization that intrinsic SHE is inherently a geometric property that arises from the Berry curvature of the bands in a material. These geometric properties are enhanced, or maximized, when the degrees of freedom of a wavefunction are freely (maximally) allowed to rotate. Here we demonstrate that such rotation of the degrees of freedom is maximized under weak crystal fields and structural distortions. Therefore, weak crystal fields and structural distortions are two important factors that control the value of SHE in transition metal oxides (TMOs). Additionally, we derive three other factors that control SHE values, namely, optimal positioning of the Fermi level, mixing of \textit{J}\textsubscript{eff} = 1/2 \& 3/2 bands, and electron correlations. In contrast to the expectation that moderate electron correlations generally enhance spin orbit coupling (SOC) by localizing electrons \cite{Krempa_correlated} and should therefore enhance SHE values, we report that the effect of correlations on SHE is more nuanced and depends on band structure details. Together, our findings present five rational design principles for the discovery and design of \textit{giant} SHE. Although derived specifically for SHE in TMOs, our design principles and insights are \textit{general} enough to apply to transition metal compounds, as well as, to other geometric properties like the anomalous Hall effect, orbital Hall effect, valley Hall effect etc. These design principles acquire enhanced potency due to the unique tunability of transition metal compounds. Transition metal compounds have multiple degrees of freedom, namely, charge, spin, orbital, and lattice, that are intertwined with one another \cite{TMO_DOF}, and constitute a vast tunable space ideal for the design and control of properties. Moreover, this deeper understanding of the factors that control SHE, could help the development of the highly attractive, \textit{in-situ}, external field control of SHE.

Implementing our design principles, we further discover that new materials (BCC-Pt\textsubscript{3}O\textsubscript{4} and 5d-anti-perovskites) promise to demonstrate \textit{giant} efficiencies of conversion of charge current to spin current, that are an order of magnitude larger than that reported for any oxide so far. We also enumerate large efficiency values found in a significant number of other oxides. Identifying materials with large conversion efficiency is critical for low-energy spintronic applications, with electrical control of magnetism \cite{Yan_Large, Zhang_Giant}. This conversion efficiency is captured by the spin Hall angle, which is defined as the ratio of the spin current to the charge current flowing in a material.  While our reported values significantly advance the cutting-edge values seen in oxides, they are smaller than the values seen in two topological insulators \cite{Mahendra_room_temp, Flatte_tunable_2015, Khang_conductive_BiSb}. However, the application of our general, rational design principles, in combination with the tunability of transition metal compounds, is likely to notably enhance these SHE values. Moreover, the estimated longitudnal conductivities for our best materials are $\geq 5 \times 10^2$ S/cm, making them suitable for applications in emergent memory technology \cite{Khang_conductive_BiSb}. Altogether, our work answers fundamental questions about the physics of SHE, it discovers exciting new spin Hall materials, it opens further avenues of materials research for giant SHE, and leads to new possibilities for the development of energy-efficient, next-generation memory devices.

\section{Results}
\subsection{General origin of the spin Hall effect}
We now derive the general factors that control the value of spin Hall effect, specifically, the type of spin Hall effect that arises from the intrinsic bandstructure of a material, called the intrinsic spin Hall effect \cite{Sinova_ISHE,Murakami_ISHE}. We start with the expression for spin Hall conductivity (SHC) obtained from linear response theory. Let $\hat{H}$ represent the total Hamiltonian of system and $\hat{H}_0$ represent the Hamiltonian without SOC ($\lambda \hat{l}\cdot\hat{s}$). Let $\sigma^{n,s}_{\alpha\beta}$ and $\Omega^{n,s}_{\alpha\beta}$ represent the $\alpha\beta$ component of the band (\textit{n}) resolved SHC and spin Berry curvature, respectively, of this system. The spin momentum is along the direction \textit{s}, and $\alpha,\beta, s$ are along the coordinate axes (\textit{x,y,z}).
\begin{equation}\label{eq1}
\hat{H} = \hat{H}_0 + \lambda \hat{l}\cdot\hat{s}
\end{equation}
\begin{equation}\label{eq2}
\sigma^s_{\alpha\beta} = -\frac{e}{\hbar}\int_{BZ}\frac{\vec{dk}}{(2\pi)^3}\sum_n f_n(\vec{k})\Omega^{n_{\vec{k}},s}_{\alpha\beta}
\end{equation}
In the above, suppressing the index $\vec{k}$:
\begin{equation}\label{eq3}
\Omega^{n,s}_{\alpha\beta} = -2 \text{Im} [\sum_{m\neq n}\frac{j^s_{nm,\alpha}v_{mn,\beta}}{(\epsilon_m-\epsilon_n)^2}]
\end{equation}
\begin{equation}\label{eq4}
j^s_{nm,\alpha} = \frac{1}{2}\sum_l[\langle n| \hat{s}_s|l \rangle \langle l|\hat{v}_\alpha |m \rangle + \langle n|\hat{v}_\alpha |l \rangle \langle l |\hat{s}_s | m \rangle]
\end{equation}
\begin{equation}\label{eq5}
v_{mn,\beta} = \langle m | \hat{v}_\beta | n \rangle
\end{equation}
These equations demonstrate that the spin Berry curvature originates from the product of the spin operator ($\hat{s}_s$) and the anomalous velocity operators ($\hat{v}_\alpha, \hat{v}_\beta$). We assume that the spin conserving part of the SOC operator ($\lambda\hat{l}_z\hat{s}_z$) makes the dominant contribution to the SHC, as compared to the spin mixing part. We include our justification for this assumption in supplementary material and show that this assumption is often, though not always, valid.
Substituting \ref{eq4} into \ref{eq3}, we obtain:
\begin{equation}\label{eq6}
\Omega^{n,s}_{\alpha\beta} = -\text{Im} \sum_{m\neq n,l}\frac{[\langle n| \hat{s}_s|l \rangle \langle l|\hat{v}_\alpha |m \rangle + \langle n|\hat{v}_\alpha |l \rangle \langle l |\hat{s}_s | m \rangle]}{(\epsilon_m-\epsilon_n)^2}.
\end{equation}
In the spin basis, we can write the time reversed pair of eigenstates $|n_{\vec{k}} \rangle$ and  $|n_{-\vec{k}} \rangle$ as:
\begin{equation}\label{eq7}
|n_{\vec{k}} \rangle \equiv  
\begin{bmatrix}
    \mu_n   \\
    \nu_n
    \end{bmatrix}
|n_{-\vec{k}} \rangle \equiv  
\begin{bmatrix}
    -\nu_n^*   \\
    \mu_n^*
    \end{bmatrix}
\end{equation}
According to our assumption, we neglect any contribution to SHC from the spin mixing parts of SOC. The expression for $\Omega^{n,s}_{\alpha\beta}$ can thus be simplified to include only the anomalous velocity terms that arise from the spin conserving part of the Hamiltonian, 
\begin{equation}\label{eq8}
\hat{v}_\alpha = \frac{\partial}{\partial k_\alpha}(\hat{H}_0 + \lambda\hat{l}_z\hat{s}_z) = \begin{bmatrix}
    \hat{v}^0_\alpha+\hat{v}^z_\alpha    &0   \\
    0       &\hat{v}^0_\alpha-\hat{v}^z_\alpha
    \end{bmatrix},
\end{equation}
where $\hat{v}^0_\alpha = \partial/\partial k_\alpha (\hat{H}_0)$ and $\hat{v}^z_\alpha = \partial/\partial k_\alpha (\lambda\hat{l}_z/2)$. If $\hat{v}_\alpha$ and $\hat{v}_\beta$ cannot mix spin, neither can $\hat{s}_s$. Thus the non-zero terms in \ref{eq6} are:
\begin{equation}\label{eq9}
\begin{split}
\langle n_{\vec{k}} | \hat{s}_s | l_{\vec{k}} \rangle = \langle \mu_n | \mu_l \rangle - \langle \nu_n | \nu_l \rangle \equiv s_{nl};
\langle n_{-\vec{k}} | \hat{s}_s | l_{-\vec{k}} \rangle = \langle \nu_l | \nu_n \rangle - \langle \mu_l | \mu_n \rangle = -s_{nl}^* \\
\langle l_{\vec{k}} | \hat{v}_\alpha | m_{\vec{k}} \rangle = \langle \mu_l |\hat{v}^0_\alpha + \hat{v}^z_\alpha| \mu_m \rangle + \langle \nu_l |\hat{v}^0_\alpha - \hat{v}^z_\alpha| \nu_m \rangle = v^0_{\alpha,lm} + v^z_{\alpha,lm}\\
\langle l_{-\vec{k}} | \hat{v}_\alpha | m_{-\vec{k}} \rangle = \langle \mu_l |\hat{v}^0_\alpha - \hat{v}^z_\alpha| \mu_m \rangle^* + \langle \nu_l |\hat{v}^0_\alpha + \hat{v}^z_\alpha| \nu_m \rangle^* = (v^0_{\alpha,lm})^* - (v^z_{\alpha,lm})^*\\
where: v^0_{\alpha,lm} \equiv \langle \mu_l |\hat{v}^0_\alpha | \mu_m \rangle + \langle \nu_l |\hat{v}^0_\alpha | \nu_m \rangle;
v^z_{\alpha,lm} \equiv \langle \mu_l |\hat{v}^z_\alpha | \mu_m \rangle - \langle \nu_l |\hat{v}^z_\alpha | \nu_m \rangle.\\
\end{split}
\end{equation}
Substituting $\alpha$ with $\beta$, \textit{n} with \textit{m} and \textit{l} with \textit{n}, we write the complete expression for the spin Berry curvature for the eigenstate $|n_{\vec{k}}\rangle$, i.e., $\Omega^{n_{\vec{k}},s}_{\alpha\beta}$ as:
\begin{equation}\label{eq10}
\Omega^{n_{\vec{k}},s}_{\alpha\beta} = -\text{Im} \sum_{m\neq n,l}\frac{[s_{nl}(v^0_{\alpha,lm} + v^z_{\alpha,lm}) + s_{lm}(v^0_{\alpha,nl} + v^z_{\alpha,nl})](v^0_{\beta,mn} + v^z_{\beta,mn})}{(\epsilon_m-\epsilon_n)^2}.
\end{equation}
Similarly, we write the spin Berry curvature for the time reversal partner of $|n_{\vec{k}}\rangle$, which is $|n_{-\vec{k}}\rangle$, i.e., $\Omega^{n_{-\vec{k}},s}_{\alpha\beta}$ as:
\begin{equation}\label{eq11}
\Omega^{n_{-\vec{k}},s}_{\alpha\beta} = -\text{Im} \sum_{m\neq n,l}\frac{[-s_{nl}^*(v^0_{\alpha,lm} - v^z_{\alpha,lm})^* - s_{lm}^*(v^0_{\alpha,nl} - v^z_{\alpha,nl})^*](v^0_{\beta,mn} - v^z_{\beta,mn})^*}{(\epsilon_m-\epsilon_n)^2}.
\end{equation}
The total SHC of the material would involve adding the spin Berry curvature contributions from both the time reversal partners. Comparing \ref{eq10} and \ref{eq11} and using $\text{Im}(A^*B^*C^*)=-\text{Im}(ABC)$, we are left with only two non-zero terms that can contribute to the total SHC ($\sigma^s_{\alpha\beta}$). The first term is $(s_{nl}v^z_{\alpha,lm}v^z_{\beta,mn} + s_{lm}v^z_{\alpha,nl}v^z_{\beta,mn})$. This term arises from the anomalous velocity originating in the rotation of eigenstates, or in other words, in the mixing of orbitals, caused by the spin conserving part of SOC. The second term $(s_{nl}v^0_{\alpha,lm}v^0_{\beta,mn} + s_{lm}v^0_{\alpha,nl}v^0_{\beta,mn})$ arises from the anomalous velocity originating in the rotation of eigenstates caused by the spin independent part of the Hamiltonian ($\hat{H}_0$). For materials with inversion and time reversal symmetry, which includes all materials in this study, the second term which is related to the Berry curvature, is zero \cite{Xiao_Berry_2010}.

In the basis of the \textit{d} orbitals of TMOs, $\hat{l}_z$ mixes either $d_{xz}$ with $d_{yz}$ orbitals or it mixes $d_{xy}$ with $d_{x^2-y^2}$ orbitals. Consequently, for $\hat{v}^z_\alpha = \partial/\partial k_\alpha (\lambda\hat{l}_z/2)$, the only non-zero components of $\hat{v}^z_{\alpha}$ arise either from the mixing of $d_{xz}$ with $d_{yz}$ orbitals or the mixing of $d_{xy}$ with $d_{x^2-y^2}$ orbitals. From the derivation above we deduce the general rules for obtaining significant spin Hall effect in materials with inversion and time reversal symmetry, as enumerated below.

\textbf{\textit{R1}}. Firstly, non-zero SHE requires the bands of character $d_i$ and $d_j$ to be close enough in energy at a given $\vec{k}$, such that the SOC can mix these states, where $i=xz, j=yz, \, or, \, i=xy, j=x^2-y^2$ for \textit{d} orbitals. 

\textbf{\textit{R2}}. Secondly, for non-zero SHC, the Fermi level should pass through the region where the bands $d_i$ and $d_j$ are crossing or being mixed by SOC. If the Fermi level does not pass through the band mixing region, all interacting states will have the same occupation. As a result, $f_n(\vec{k})v^z_{\alpha,nm} v^z_{\beta,mn}$ will exactly cancel with $f_n(\vec{k})v^z_{\alpha,mn} v^z_{\beta,nm}$ as $l_{z,mn} = l_{z,nm}^*$. Thus, the Fermi level positioning is critical to the total value of SHC. 

\textbf{\textit{R3}}. Thirdly, the coupling of states via SOC will contribute to spin Berry curvature only if there are net transitions between $d_i$ and $d_j$.

\subsection{Rational design principles for large SHE in TMOs}
\begin{figure}
    \centering
    \includegraphics[scale=0.75]{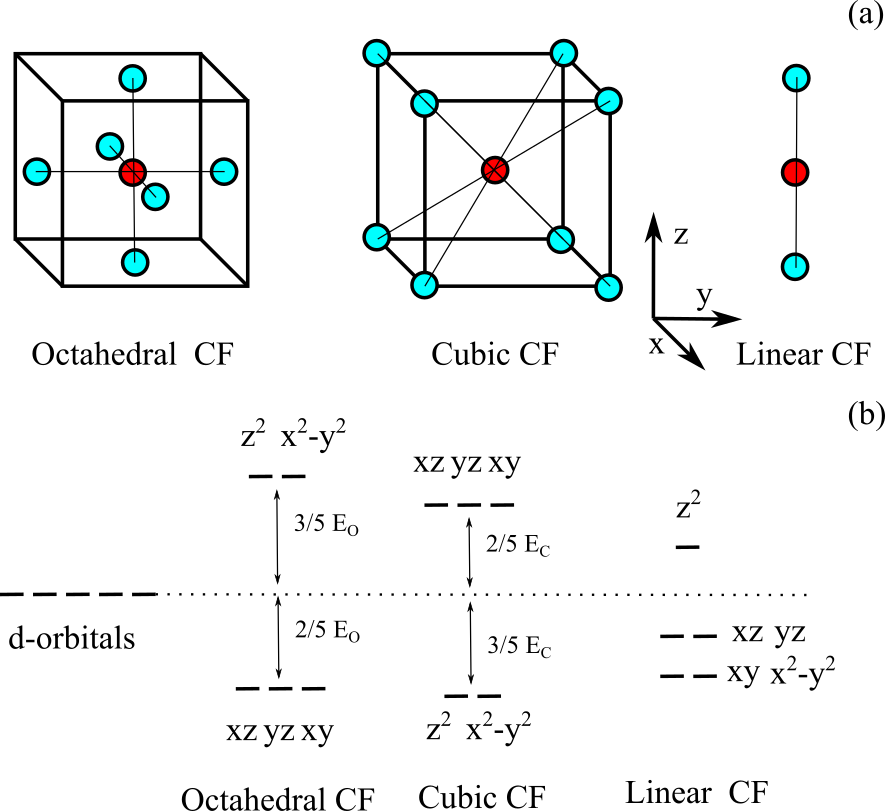}
    \caption{Illustration of crystal field splitting in transition metal oxides. Section (a) demonstrates structural diagrams of octahedral, cubic and linear crystal fields (CFs), with the transition metal atom marked in red and the ligand atom marked in cyan. Section (b) plots the energy splittings of \textit{d} orbitals generated under these crystal fields.}
    \label{fig:1}
\end{figure}

In the section above we have derived three basic rules that govern the value of SHC in \textit{d} orbital materials. From these we will now deduce the rational design principles for large SHC in TMOs. The band structure of a transition metal oxide is generally explained with the help of crystal field theory (CFT) \cite{Bethe_Termaufspaltung, Vleck_Theory}. Crystal field theory considers a transition metal atom with its detailed electronic structure in an environment of ligands that are considered structure-less. The primary effect of the ligands is to alter the electronic structure of the transition metal due to repulsion between the transition metal electrons and the electronegative ligand ions. The repulsive ligand field, and consequently the electronic structure, is determined by the geometry of the ligands surrounding the transition metal. For instance, in most TMOs considered here, the transition metal is surrounded by six oxygen atoms that form an octahedra, resulting in what is called an octahedral crystal field (see Fig. \ref{fig:1}). For an octahedral crystal field, amongst the \textit{5d} orbitals of the transition metal, $d_{x^2-y^2}$ and $d_{z^2}$ orbitals point in the direction of the ligands and experience strong repulsion. These orbitals are thus pushed higher in energy in the bandstructure and together constitute what is called the $e_g$ manifold. On the other hand, the $d_{xz}$,\, $d_{yz}$ and $d_{xy}$ orbitals point away from the repulsive ligands, thereby experiencing lower repulsion and lying lower in energy in the bandstructure. These latter three orbitals constitute the $t_{2g}$ manifold. The concerted effect of an octahedral crystal field on the electronic structure is to split the otherwise degenerate \textit{5d}-bands of the transition metal into a lower energy three-fold degenerate $t_{2g}$ and an higher energy two-fold degenerate $e_g$ (see Fig. \ref{fig:1}). In contrast, a cubic crystal field comprises of a transition metal surrounded by eight ligand ions arranged at the corners of a cube. This cubic field splits the transition metal \textit{5d}-bands into an inverse, lower energy $e_g$ and a higher energy $t_{2g}$ manifold.

 Putting together CFT with the general rules (\textit{R}) controlling the value of SHC that we had derived before, we will now deduce our five rational design principles to obtain large SHE. Our first principle originates in the inherently geometric nature of the Berry curvature. As a geometric property, Berry curvature results from the rotation of a degree of freedom (DOF) of the electronic wavefunction with a variation of some periodic parameter \cite{Xiao_Berry_2010}. For SHE, which derives from the spin Berry curvature, the relevant degree of freedom is the orbital angular momentum (\textit{l}) and the relevant parameter is the wavevector ($\vec{k}$). For \textit{d}-orbitals, there are five possible orbital degrees of freedom (DOFs). However in TMOs, the crystal field splits this orbital space into smaller subspaces spanned by fewer DOFs. For instance, in octahedral and cubic fields the \textit{d}-bands are split into an $e_g$ (DOF=2) and a $t_{2g}$ (DOF=3) manifold. The splitting of the larger manifold into these smaller spaces reduces the possibilities for rotation of DOFs, thereby reducing the possible spin Berry curvature, and consequently reducing SHE. Therefore, we deduce that large values of SHE are encouraged by \textbf{weak crystal fields}. This is \textbf{condition 1 (\textit{C1})}. Specifically for TMOs, we have shown above that there are two possible contributions to SHE, one that arises from the mixing of $d_{xz}$ and $d_{yz}$ orbitals, and another that arises from the mixing of $d_{xy}$ and $d_{x^2-y^2}$ orbitals. It follows that SHE is maximized when both these contributions are non-zero and add constructively to one another. From rule \textit{R1}, for these two SHE contributions to be non-zero, we would require the $d_{xz}$ \& $d_{yz}$ bands, as well as, the $d_{xy}$ and $d_{x^2-y^2}$ bands to cross one another or be close enough in energy to enable SOC to mix them. The splitting of two \textit{d} bands by crystal field into separate manifolds will hamper the mixing of those two bands. As such, the first principle to obtain large values of SHE (\textit{C1}) is that the crystal field should be weak. For instance, in a weak octahedral or cubic crystal field, the $e_g$ and $t_{2g}$ manifolds overlap allowing band crossings between the $d_{xy}$ and $d_{x^2-y^2}$ orbitals. 
 
 In materials with strong crystal fields, mixing between otherwise separate manifolds can be enhanced if structural distortions are present. Structural distortions of octahedral crystal fields, such as octahedral rotation and tilting, have been shown to break up the $e_g$ and $t_{2g}$ manifolds into further sub-bands for SrIrO\textsubscript{3} \cite{Nie_interplay_2015}. This splitting up of a manifold increases the total bandwidth of that manifold, which encourages overlaps or transitions between separate manifolds. In other words, structural distortions increase the space of rotations of the orbital degrees of freedom. We would thus expect the presence of structural distortions to mitigate the dampening effect of a strong crystal field on SHE, and generate large SHE values. We call this condition of \textbf{structural distortions}, \textbf{condition 2 (\textit{C2})}. The enhancement of SHE via distortions has been experimentally reported for SrIrO\textsubscript{3} \cite{Nan_anisotropic_SrIrO3}. Although, we have derived conditions 1 and 2 (\textit{C1 \& C2}) specifically for octahedral crystal fields, they are general conditions extending beyond octahedral or even cubic crystal fields. This is because, in general, the weaker the crystal field, and the greater the structural distortions, the greater the possibility of mixing between various \textit{d}-orbitals. Condition 3 (\textit{C3}) is the same as the general rule 2 (\textit{R2}), i.e., the \textbf{optimal positioning of the Fermi level}.
 
  To deduce condition 4 (\textit{C4}), we only focus on the $t_{2g}$ manifold of octahedral and cubic crystal fields, and the transitions therein between the $d_{xz}$ and $d_{yz}$ orbitals. The impact of SOC is known to split the $t_{2g}$ manifold into a lower energy \textit{J}\textsubscript{eff} = 3/2 and a higher energy \textit{J}\textsubscript{eff} = 1/2 sub-manifold \cite{Stamokostas_mixing}. The electron wavefunctions for the states in these \textit{J}\textsubscript{eff} = 3/2 \& 1/2 sub-manifolds can be written as:
 \begin{equation}\label{eq20}
 \begin{split}
|j_\text{eff}=3/2,m_j=3/2 \rangle= -\frac{1}{\sqrt{2}}d_{yz}\otimes|\uparrow \rangle - \frac{i}{\sqrt{2}}d_{xz}\otimes|\uparrow \rangle \\
|j_\text{eff}=3/2,m_j=1/2 \rangle= -\frac{1}{\sqrt{6}}d_{yz}\otimes|\downarrow \rangle - \frac{i}{\sqrt{6}}d_{xz}\otimes|\downarrow \rangle +\sqrt{\frac{2}{3}}d_{xy}\otimes|\uparrow \rangle \\
|j_\text{eff}=3/2,m_j=-1/2 \rangle= \frac{1}{\sqrt{6}}d_{yz}\otimes|\uparrow \rangle - \frac{i}{\sqrt{6}}d_{xz}\otimes|\uparrow \rangle +\sqrt{\frac{2}{3}}d_{xy}\otimes|\downarrow \rangle \\
|j_\text{eff}=3/2,m_j=-3/2 \rangle= \frac{1}{\sqrt{2}}d_{yz}\otimes|\downarrow \rangle - \frac{i}{\sqrt{2}}d_{xz}\otimes|\downarrow \rangle
\end{split}
\end{equation}
 \begin{equation}\label{eq21}
 \begin{split}
|j_\text{eff}=1/2,m_j=1/2 \rangle= \frac{1}{\sqrt{3}}d_{yz}\otimes|\downarrow \rangle + \frac{i}{\sqrt{3}}d_{xz}\otimes|\downarrow \rangle +\sqrt{\frac{1}{3}}d_{xy}\otimes|\uparrow \rangle \\
|j_\text{eff}=1/2,m_j=-1/2 \rangle= \frac{1}{\sqrt{3}}d_{yz}\otimes|\uparrow \rangle - \frac{i}{\sqrt{3}}d_{xz}\otimes|\uparrow \rangle -\sqrt{\frac{1}{3}}d_{xy}\otimes|\downarrow \rangle
\end{split}
\end{equation}
where $|\uparrow \rangle \, \& |\downarrow \rangle$ are spin up and spin down states, respectively. In octahedral and cubic crystal fields, before the action of SOC, if the $d_{xz}$ and $d_{yz}$ orbitals are exactly degenerate, the electron wavefunctions will have exactly symmetric contributions from these two orbitals. Once SOC is turned on, in the absence of any mixing between the \textit{J}\textsubscript{eff} = 3/2 \& 1/2 states, the spin orbit coupled wavefunctions, given by eqs. \ref{eq20}, \ref{eq21}, will also have exactly symmetric contributions from $d_{xz}$ and $d_{yz}$ orbitals. In such a case, the action of SOC would not lead to any $d_{xz} - d_{yz}$ transitions, and no resulting contribution to SHC (refer to rule \textit{R3}). The reverse condition, i.e., octahedral and cubic crystal fields with degenerate $d_{xz}$ and $d_{yz}$ orbitals along with a mixing of \textit{J}\textsubscript{eff} = 3/2 \& 1/2 states could result in enhanced SHE values. We call this \textbf{condition 4 (\textit{C4})}. It is interesting to note that one way to enhance the mixing of \textit{J}\textsubscript{eff} = 3/2 \& 1/2 sub-manifolds is via structural distortions \cite{Nie_interplay_2015}, thereby enhancing $d_{xz}-d_{yz}$ transitions. However, under CFT, structural distortions can themselves break the degeneracy between $d_{xz}$ and $d_{yz}$ orbitals, also generating $d_{xz}-d_{yz}$ transitions. Whether the two effects cooperate or cancel is not immediately obvious and would likely depend on the band structure details. Finally, \textbf{condition 5 (\textit{C5})} is that the presence of \textbf{moderate electron correlation} can enhance SHE by localizing electrons, as has been discussed in literature \cite{Krempa_correlated}. However, later we show that this effect is nuanced and depends on band structure details. 

To summarize the five general rational design principles for large SHE are:

\textbf{\textit{C1}}. Weak crystal fields.

\textbf{\textit{C2}}. Structural distortions.

\textbf{\textit{C3}}. Optimal positioning of the Fermi level.

\textbf{\textit{C4}}. Mixing of \textit{J}\textsubscript{eff} = 1/2 \& 3/2 bands, for degenerate $d_{xz}$ and $d_{yz}$ orbitals, under octahedral or cubic crystal fields.

\textbf{\textit{C5}}. Moderate electron correlations.

In the remaining sections, we report our findings on SHE in various \textit{5d}-TMOs and explain them on the basis of these five rational design principles.

\subsection{The effect of weak crystal field (\textit{C1})}
We first report the finding of giant SHE in BCC-Pt\textsubscript{3}O\textsubscript{4}. Pt\textsubscript{3}O\textsubscript{4} has two possible proposed crystal structures \cite{Seriani_Catalytic}, body centered cubic (BCC) \cite{Galloni_Pt3O4} and simple cubic (SC) \cite{Muller_Pt3O4}. BCC-Pt\textsubscript{3}O\textsubscript{4} consists of Pt atoms surrounded by a cubic oxygen crystal field, while SC-Pt\textsubscript{3}O\textsubscript{4} has Pt atoms inside a square planar crystal field. As we show below, the crystal field in BCC-Pt\textsubscript{3}O\textsubscript{4} is weak which satisfies condition \textit{C1}, leading to a giant SHE. The calculated spin Hall conductivity (SHC) for BCC-Pt\textsubscript{3}O\textsubscript{4} is $3.7 \times 10^3 \, \hbar/2e \, S/cm$ with an estimated giant spin Hall angle $\Theta$\textsubscript{SH} of $\sim 7.4$. In sharp contrast to BCC-Pt\textsubscript{3}O\textsubscript{4}, SC-Pt\textsubscript{3}O\textsubscript{4} displays a small SHC as a result of its strong crystal field which violates condition \textit{C1}. Another exciting result is the finding of a hypothetically giant SHE in the rare earth anti-perovskite Yb\textsubscript{3}PbO. Yb\textsubscript{3}PbO has Yb in a weak crystal field, satisfying condition \textit{C1}. However, unfortunately, the nominal Fermi level in Yb\textsubscript{3}PbO is poorly placed which violates condition \textit{C3}. Engineering the Fermi level to place it at 0.6 eV above the nominal Fermi level of Yb\textsubscript{3}PbO satisfies the condition \textit{C3}, leading to an SHC of $-2.4 \times 10^3 \, \hbar/2e \, S/cm$ with an estimated giant $\Theta$\textsubscript{SH} of $-2.5 \, \text{to} \, -4.9$. Though hypothetical, this latter result, alerts us to the possibility of finding giant SHE in \textit{5d}-transition metal anti-perovskites via careful engineering of the Fermi level. Our preliminary results for SHE in TM\textsubscript{3}PbO, where Yb atoms have been substituted by \textit{5d}-transition metal (TM) atoms Hf and Ta, show promise towards achieving giant SHE in anti-perovskites. We will report these results in detail, in future work. The giant values of $\Theta$\textsubscript{SH} described in this section (and in Table I), are at least an order of magnitude larger than those reported for any TMOs so far. Additionally, these values are greater than those observed for all other measured materials except for $\text{Bi}_{\text{x}}\text{Se}_\text{1-x}$ and $\text{Bi}_{\text{1-x}}\text{Sb}_\text{x}$. Our spin Hall results give further support to the rational design principles deduced above.

\begin{figure}
    \centering
    \includegraphics[scale=0.70]{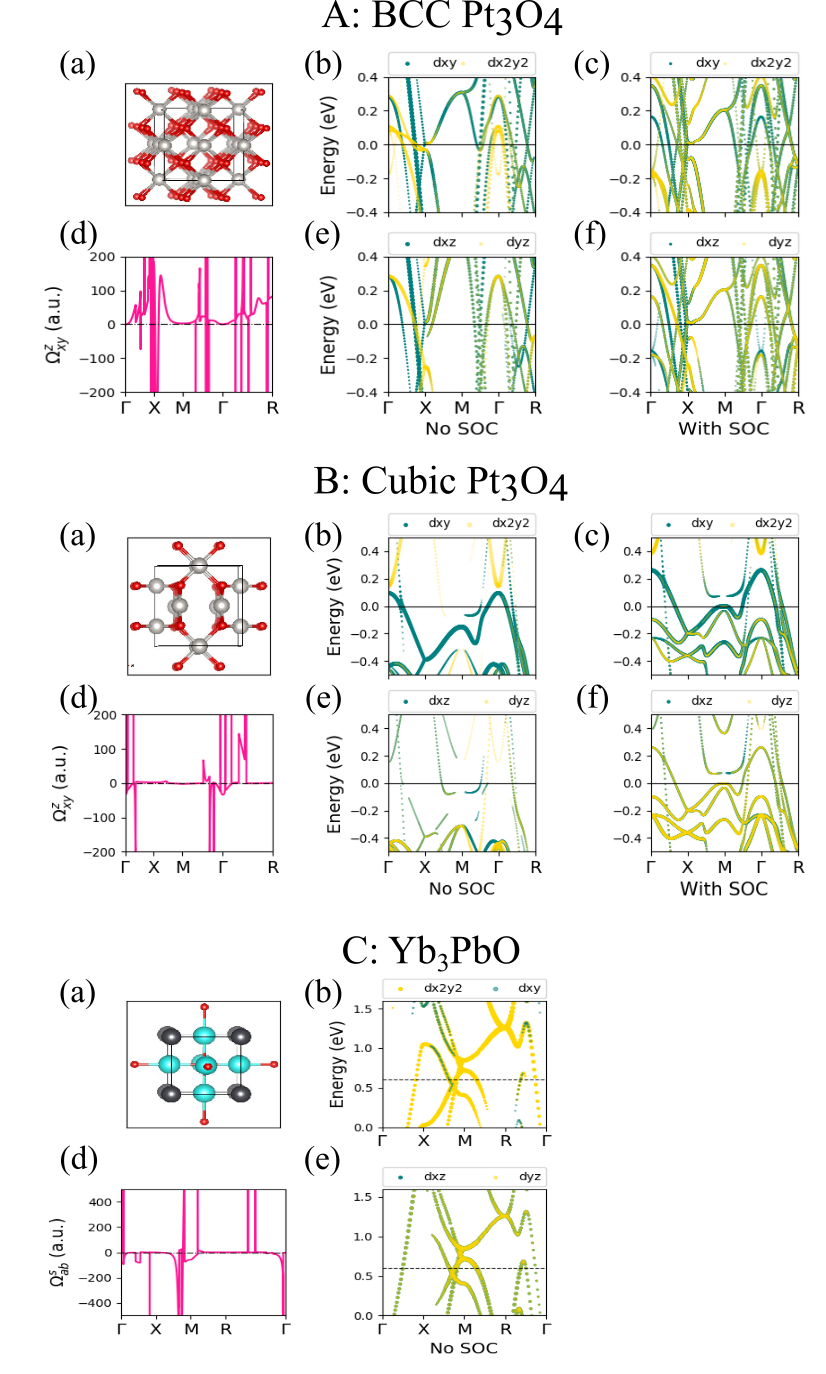}
    \caption{SHE in BCC-Pt\textsubscript{3}O\textsubscript{4} (panel A), SC-Pt\textsubscript{3}O\textsubscript{4} (panel B) and Yb\textsubscript{3}PbO (panel C). Section (a) displays the structure of these oxides with Pt in silver, Yb in cyan, Pb in dark gray and O in red, (b) \& (c) show the bandstructure projected onto $d_{xy}$ orbitals (green) and $d_{x^2-y^2}$ orbitals (yellow), without and with SOC, respectively, (e) \& (f) plot the bandstructure projected onto $d_{xz}$ orbitals (green) and $d_{yz}$ orbitals (yellow), without and with SOC, respectively, and (d) portrays the spin Berry curvature.}
    \label{fig:2}
\end{figure}

BCC-Pt\textsubscript{3}O\textsubscript{4} has a cubic crystal field which splits the \textit{5d}-bands into a lower energy $e_g$ and a higher energy $t_{2g}$ manifold. A large Pt-O bond length of 2.7 \AA \, leads to a weak crystal field (\textit{C1}), which is evident from a significant overlap between the $d_{xy}$ orbitals of the $e_g$ manifold and the $d_{x^2-y^2}$ orbitals of the $t_{2g}$ manifold, as seen in Fig. \ref{fig:2}A:b,c. This overlap allows for multiple band crossings between the lower $d_{x^2-y^2}$ and the upper $d_{xy}$ bands, along $\Gamma$-X as well as $\Gamma$-R in k-space, at $E_F$. Additionally, $d_{xz}$ and $d_{yz}$ bands also demonstrate various band crossings in this k-space region (see Fig. \ref{fig:2}A:e). Turning on SOC leads to a clear mixing of these bands and a splitting of degeneracies (see Fig. \ref{fig:2}A:c,f), creating spin orbit coupled states. This band mixing generates large spin Berry curvature along $\Gamma$-X and $\Gamma$-R, as shown in Fig. \ref{fig:2}A:d, which results in a giant SHC. Note that the Fermi level is well-positioned as it passes through the region of band mixing. On the other hand, SC-Pt\textsubscript{3}O\textsubscript{4} has Pt in a strong square planar crystal field, which violates condition \textit{C1} leading to a significantly lower estimated $\Theta$\textsubscript{SH} of 0.5. A square planar crystal field splits the \textit{5d}-bands into multiple manifolds, starting from a higher energy $d_{x^2-y^2}$ level, followed by $d_{xy}$ and even lower $d_{z^2}$ levels, and ending with the lowest energy level comprised of degenerate $d_{xz}$ and $d_{yz}$ bands. The strength of the square planar crystal field in SC-Pt\textsubscript{3}O\textsubscript{4} is evident from the insignificant overlap between the $d_{xy}$ and $d_{x^2-y^2}$ orbitals, as shown in Fig. \ref{fig:2}B:b. Additionally, in the absence of SOC, we observe a minimal presence of $d_{xz}$ and $d_{yz}$ bands at the Fermi level, all of which leads to a much smaller SHC than that seen for BCC-Pt\textsubscript{3}O\textsubscript{4}. 

Yb\textsubscript{3}PbO is an anti-perovskite, where the A and B sites of a regular perovskite have been swapped such that the central atom (Yb) is on the A site and Pb is on the B site. Every Yb atom is bonded to only two oxygen atoms with an O-Yb-O bond angle of 180\textdegree, constituting a linear ligand crystal field. Comprising only two ligand atoms makes the linear crystal field weak, satisfying condition \text{C1} and allowing for a large SHE. This crystal field splits the \textit{5d} bands into smaller manifolds, comprising degenerate $d_{xz}$ and $d_{yz}$ bands, and separately degenerate $d_{xy}$ and $d_{x^2-y^2}$ bands. The weakness of this linear crystal field is evident from the overlap between these manifolds as shown in Fig. \ref{fig:2}C:b,e. The degeneracy between $d_{xy}$ and $d_{x^2-y^2}$ orbitals is further broken by a weak repulsion of the Yb-$d_{xy}$ orbitals from the B site Pb ions (see Fig. \ref{fig:2}C:a) that pushes the $d_{xy}$ bands higher in energy (see Fig. \ref{fig:2}C:b). The concerted effort of the weak ligand crystal field (condition \textit{C1}), and a weak repulsive field from the cation Pb (modified condition \textit{C1}), along with a hypothetical increase in the Fermi level by 0.6 eV (condition \textit{C3}), leads to a giant SHE. At 0.6 eV above $E_F$, close to the \textit{M} point, there is a band crossing between the lower Yb-$d_{x^2-y^2}$ bands and the upper Yb-$d_{xy}$ bands, along with the presence of $d_{xz} \, \& \, d_{yz}$ orbitals (see Fig. \ref{fig:2}C:b, e). SOC is expected to mix these bands and create a large hot spot for spin Berry curvature around \textit{M}, which is shown in Fig. \ref{fig:2}C:d. As a result, we see a giant estimated SH angle in Yb\textsubscript{3}PbO for a Fermi level positioning of 0.6 eV above the nominal value. Yb\textsubscript{3}PbO is identified as a possible topological crystalline insulator \cite{Zhang_Catalogue} with a Dirac node at the $E_F$ \cite{Pertsova_computational}, however we do not see any significant SHE contribution from this node.

Here we would like to note that while technically Yb\textsubscript{3}PbO is a rare-earth oxide and not a TMO, a Fermi level which is 0.6 eV above the nominal Fermi level of Yb\textsubscript{3}PbO passes through the empty \textit{5d} bands of Yb, which brings our design principles into effect. As mentioned before, the optimal positioning of the Fermi level in anti-perovskites for giant SHE, can be achieved by substituting Yb atoms with the 5d-transition metal atoms Hf and Ta, such that the Fermi level passes through the \textit{5d} bands. We also note that there exist many known 5d-transition metal anti-perovskite compounds (beyond oxides) such as Pt\textsubscript{3}REB (RE: Rare Earth) \cite{Mondal_Structural}, Pt\textsubscript{3}AP (A = Sr, Ca, La) \cite{Pt3AP}, while many more are theoretically predicted to be stable \cite{Bannikov_Elastic} and remain to be synthesized. Our preliminary calculations on these transition metal anti-perovskites show that giant spin Hall values can be obtained in anti-perovskites by careful engineering of the materials' chemical composition. We will report these results in detail in future work.

\begin{table}
\centering
\begin{tabular}{||c||c||c||c||}
 \hline \hline
 \multicolumn{4}{||c||}{Table I: List of SHC $\sigma^s_{\alpha\beta}$ ($\hbar/2e \, S/cm$), longitudnal conductivity $\sigma$ and spin Hall angle $\Theta^s_{\alpha\beta}$}\\
 \hline
 Structure& $\sigma^z_{xy}$=$\sigma^x_{yz}$=$\sigma^y_{zx}$&$\sigma$ (S/cm)&$\Theta^z_{xy}$=$\Theta^x_{yz}$=$\Theta^y_{zx}$ \\
 \hline \hline
\multicolumn{4}{||c||}{Experimental spin Hall values from literature for comparision}\\
\hline $\beta-$W \cite{Pai_tungsten}&       1000&	$3.3 \times 10^3$&		0.3\\
\hline Bi\textsubscript{1-x}Sb\textsubscript{x} \cite{Khang_conductive_BiSb}&   $1.3 \times 10^5$&   	$2.5 \times 10^3$&		    52\\
\hline SrIrO\textsubscript{3} \cite{Nan_anisotropic_SrIrO3}&      $1300$&	    	$2.5 \times 10^3$&		    0.5\\
\hline IrO\textsubscript{2} \cite{Fujiwara_IrO2_2013}&        -&	    $5.0 \times 10^3$&		    0.04\\
\hline \hline
\multicolumn{4}{||c||}{Prediction of giant spin Hall effect}\\
\hline Pt\textsubscript{3}O\textsubscript{4} (BCC)&     3676&  $\sim 5 \times 10^2$ \cite{Neff_1996}&  $\sim 7.4$\\
\hline Pt\textsubscript{3}O\textsubscript{4} (Cubic)&   244&  $\sim 5 \times 10^2$ \cite{Neff_1996}&  $\sim 0.5$\\
\hline Yb\textsubscript{3}PbO ($E_F$)&     -124&  $\sim 5 \times 10^2 - 10^3$ \cite{Samal_molecular}&  $\sim -0.1 \, \text{to} \, -0.3$\\
\hline Yb\textsubscript{3}PbO ($E_F$+0.6 eV)&   -2448&  $\sim 5 \times 10^2 - 10^3$ \cite{Samal_molecular}&  $\sim -2.5 \, \text{to} \, -4.9$\\
\hline \hline
\end{tabular}
\label{table:1}
\end{table}

\subsection{The role of structural distortions (\textit{C2})} 
We now examine cubic perovskites where the transition metal atom is under a typically strong, octahedral crystal field. BaOsO\textsubscript{3} \cite{Jung_electronic_BaOsO3, Ali_theoretical_AOsO3} and SrOsO\textsubscript{3} \cite{Ali_theoretical_AOsO3} are both perovskite osmates with different sizes of the A site cation. As a result, BaOsO\textsubscript{3} adopts a perfect perovskite structure free from distortions, and demonstrates a low SHC, on account of its strong crystal field violating condition 1 (\textit{C1}). In contrast, SrOsO\textsubscript{3} demonstrates large octahedral distortions, satisfying condition 2 (\textit{C2}), which leads to a larger SHC for SrOsO\textsubscript{3}. To account for correlation effects in BaOsO\textsubscript{3} and SrOsO\textsubscript{3}, we use an LDA+U scheme with U = 2 eV which is taken from \cite{Ali_theoretical_AOsO3}. Like SrOsO\textsubscript{3}, SrIrO\textsubscript{3} \cite{Nie_interplay_2015} also demonstrates significant distortions of the O6 octahedra, which satisfies condition \textit{C2}, imparting a larger SHC to SrIrO\textsubscript{3} than BaOsO\textsubscript{3}. The enhancement of SHC by structural distortions has been experimentally shown for SrIrO\textsubscript{3} \cite{Nan_anisotropic_SrIrO3}. The values of SHC and $\Theta$\textsubscript{SH} for these materials are calculated along the pseudo-cubic axes and are enumerated in Table II. An analysis of SHE in the rutile binary oxides OsO\textsubscript{2} and IrO\textsubscript{2} is included in the supplementary.

\begin{figure}
    \centering
    \includegraphics[scale=0.70]{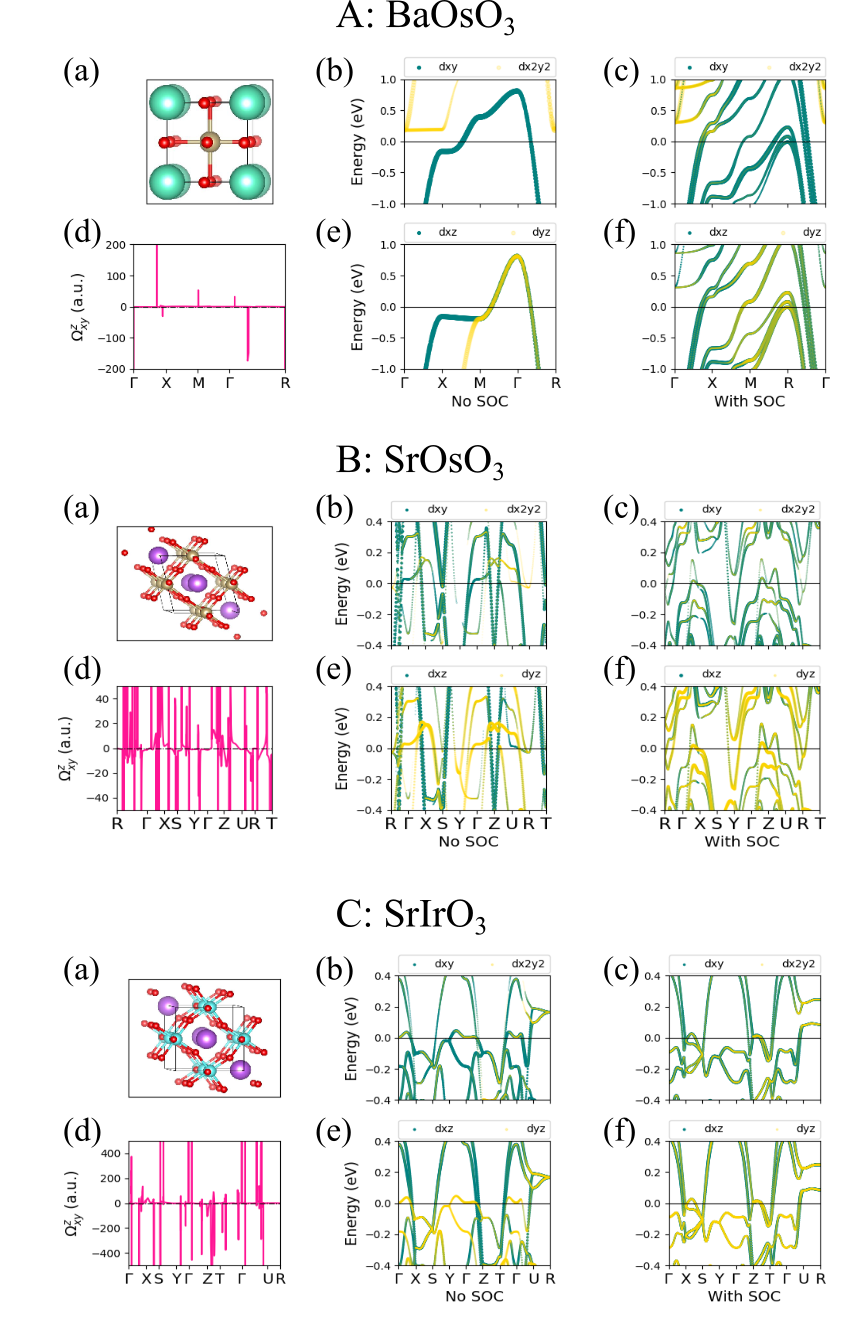}
    \caption{SHE in perovskite oxides including BaOsO\textsubscript{3} (panel A), SrOsO\textsubscript{3} (panel B) and SrIrO\textsubscript{3} (panel C). Section (a) displays the structure of these oxides with Os in yellow, Ba in green, Sr in purple, Ir in cyan and O in red, (b) \& (c) show the bandstructure projected onto $d_{xy}$ orbitals (green) and $d_{x^2-y^2}$ orbitals (yellow), without and with SOC, respectively, (e) \& (f) plot the bandstructure projected onto $d_{xz}$ orbitals (green) and $d_{yz}$ orbitals (yellow), without and with SOC, respectively, and (d) portrays the spin Berry curvature.}
    \label{fig:3}
\end{figure}

BaOsO\textsubscript{3} is a distortion-free cubic perovskite, with a metallic conductivity \cite{Jung_electronic_BaOsO3, Ali_theoretical_AOsO3}, and an octahedral crystal field that splits the \textit{5d}-bands into a lower $t_{2g}$ and an upper $e_g$ manifold, with the Fermi level lying inside the former. The strong crystal field precludes any $e_g - t_{2g}$ overlap, violating condition \textit{C1}, as is evident from Fig. \ref{fig:3}A:b and resulting in a small SHC (see Table II). 

SrOsO\textsubscript{3} \cite{Ali_theoretical_AOsO3} and SrIrO\textsubscript{3} \cite{Nie_interplay_2015} are orthorhombic perovskites with a strong octahedral crystal field (\textit{C1}), as well as the presence of structural distortions (\textit{C2}). If acting alone, a strong crystal field violates condition \textit{C1}, leading to a low SHC, as seen in BaOsO\textsubscript{3}. However, structural distortions satisfy condition \textit{C2} and mitigate the effect of crystal field, resulting in a larger SHC for SrOsO\textsubscript{3} and SrIrO\textsubscript{3} (see Table II). The mechanism by which structural distortions enhance SHE is evident from a comparision of the projected bandstructures for BaOsO\textsubscript{3} (see Fig. \ref{fig:3}A: b,c), with that for SrOsO\textsubscript{3} (see Fig. \ref{fig:3}B: b,c), or SrIrO\textsubscript{3} (see Fig. \ref{fig:3}C: b,c). Distortions split the $t_{2g}$ and $e_g$ manifolds into multiple sub-bands, increasing the energy-span of both manifolds. This splitting, in turn, increases the likelihood of crossings between different sub-bands within a manifold, as well as crossings between different manifolds. This mechanism underpins condition \textit{C2}. Although it has been shown for SrIrO\textsubscript{3} \cite{Nie_interplay_2015} that distortions also enhance the mixing of \textit{J}\textsubscript{eff} = 1/2, 3/2 states, whether this mixing contributes to SHE is difficult to ascertain, as distortions simultaneously break the degeneracy between $d_{xz}$ and $d_{yz}$ orbitals. Therefore, we are unsure if condition \textit{C4} applies.
 
The busy plots of band structure and spin Berry curvature for SrOsO\textsubscript{3} (see Fig. \ref{fig:3}B: b-f) and SrIrO\textsubscript{3} (see Fig. \ref{fig:3}C: b-f) make it difficult to isolate the exact regions of spin Hall generation. However, a broad inspection reveals multiple degeneracy splittings by the action of SOC. Inspection of projected band structures without SOC at the Fermi level reveals the presence of $e_g$ - $t_{2g}$ overlap that is greater than that seen in BaOsO\textsubscript{3}. The $e_g$ - $t_{2g}$ manifold overlap is greater for SrIrO\textsubscript{3} than SrOsO\textsubscript{3} owing to the increased band filling of the former. This overlap likely explains the order of SHC values seen which are largest for SrIrO\textsubscript{3}, followed by SrOsO\textsubscript{3} and smallest for BaOsO\textsubscript{3}. Our estimated values of $\Theta$\textsubscript{SH} for SrIrO\textsubscript{3} are $\Theta^z_{xy} \sim 0.2$ \& $\Theta^x_{yz} \sim 0.34$ (see Table II) which are comparable to the experimentally observed values of $\Theta^z_{xy} \sim 0.5 \,\, \& \,\, \Theta^x_{yz} \sim 0.3$ \cite{Nan_anisotropic_SrIrO3}.

\begin{table}\label{table:2}
\centering
\begin{tabular}{ ||p{3cm}||p{1.5cm}|p{1.5cm}|p{1.5cm}||p{2cm}||p{1.5cm}|p{1.5cm}|p{1.5cm}||}
 \hline \hline
 \multicolumn{8}{||c||}{Table II: List of SHC $\sigma^s_{\alpha\beta}$ ($\hbar/2e \, S/cm$), longitudnal conductivity $\sigma$ and spin Hall angle $\Theta^s_{\alpha\beta}$}\\
 \hline \hline
 \multicolumn{8}{||c||}{Spin Hall predictions for cubic perovskites}\\
 \hline 
 Structure& $\sigma^z_{xy}$&$\sigma^x_{yz}$&$\sigma^y_{zx}$&$\sigma$ (S/cm)&$\Theta^z_{xy}$& $\Theta^x_{yz}$& $\Theta^y_{zx}$ \\
\hline BaOsO\textsubscript{3}&    -150& -134&	-150&	55 \cite{Shi_High_pressure}&	    -2.7&	-2.4&	-2.7\\
\hline SrOsO\textsubscript{3}&    -294& 30&    276&	    128 \cite{Shi_High_pressure}&	-2.3&	0.23&	2.15\\
\hline SrIrO\textsubscript{3}&     382&	680&	84&	$2 \times 10^3$ \cite{Nie_interplay_2015}&	  0.2&	0.34&	0.05\\
\hline 
%\multicolumn{8}{||c||}{Table III: List of SHC $\sigma^s_{\alpha\beta}$ ($\hbar/2e \, S/cm$), longitudnal conductivity $\sigma$ and spin Hall angle $\Theta^s_{\alpha\beta}$}\\
 \hline
 \multicolumn{8}{||c||}{Spin Hall prediction for Tb\textsubscript{2}Ir\textsubscript{2}O\textsubscript{7}}\\
 \hline
 Correlation& $\sigma^z_{xy}$&$\sigma^x_{yz}$&$\sigma^y_{zx}$&$\sigma$ (S/cm)&$\Theta^z_{xy}$&$\Theta^x_{yz}$ &$\Theta^y_{zx}$\\
\hline U = 0.0 eV&   -111&	44&	    -142&    & & &\\
\hline U = 0.5 eV&   186&	196&	-188&    & & &\\
\hline U = 1.0 eV&   16&	15&	    -527&   & & &\\
\hline U = 1.5 eV&   284&	-10&	-136&    & & &\\
\hline U = 2.0 eV&   -374&	528&	-56&    & & &\\
\hline \hline 
\multicolumn{8}{||c||}{Spin Hall prediction for Bi\textsubscript{2}Ir\textsubscript{2}O\textsubscript{7}}\\
\hline
Correlation& $\sigma^z_{xy}$&$\sigma^x_{yz}$&$\sigma^y_{zx}$&$\sigma$ (S/cm)&$\Theta^z_{xy}$&$\Theta^x_{yz}$ &$\Theta^y_{zx}$\\
\hline U = 0.0 eV&   -122&	-512&	-158&	$\sim 714$ \cite{Lee_infrared_Bi2Ir2O7}&	    -0.2&	-0.7&	-0.2\\
\hline \hline
\end{tabular}
\end{table}

\subsection{The effect of electron correlations (\textit{C5})}
In this section, we focus on the relationship between electron correlations and spin Hall effect captured in condition 5 (\textit{C5}). It is generally expected that moderate electron correlations enhance SOC by localizing electrons \cite{Krempa_correlated}, and should therefore enhance SHE as well. However, here we find that the effect of correlations on SHE is more nuanced and depends on band structure details (\textit{C5}). While it is largely true that an increase in electron correlations increases the spin Berry curvature, the net effect of this enhancement on SHE can be complicated, due to the presence of competing spin Berry curvature hot spots, as well as local changes in band occupation caused by electron correlations. 

Pyrochlore iridates, in particular rare-earth (RE) pyrochlore iridates, have generated keen interest for their interesting properties resulting from the interaction of moderate correlation and SOC \cite{Pesin_Mott}. These properties include the anomalous hall effect \cite{Machida_unconventional}, spin-liquid state \cite{Nakatsuji_Metallic}, Weyl semimetal \cite{Wan_topological, Krempa_pyrochlore} etc. Here we study the SHE in a RE pyrochlore iridate, Tb\textsubscript{2}Ir\textsubscript{2}O\textsubscript{7} \cite{Lefrancois_anisotropy, Ishii_first}. In contrast, the pyrochlore iridate Bi\textsubscript{2}Ir\textsubscript{2}O\textsubscript{7} \cite{Qi_strong_Bi2Ir2O7, Lee_infrared_Bi2Ir2O7, Wang_experimental_Bi2Ir2O7} has low electron correlations due to strong hybridization between Ir \textit{5d} and Bi \textit{6p} electrons \cite{Qi_strong_Bi2Ir2O7}. We include electron correlations for Tb\textsubscript{2}Ir\textsubscript{2}O\textsubscript{7} under an LDA+U scheme with U ranging from 0 to 2.5 eV. Our results for SHE in Tb\textsubscript{2}Ir\textsubscript{2}O\textsubscript{7} and Bi\textsubscript{2}Ir\textsubscript{2}O\textsubscript{7} are listed in Table II, with the pyrochlore lattice vectors $\vec{a}, \vec{b} \, \& \, \vec{c}$ defined along $\hat{x}+\hat{y}$, $\hat{x}-\hat{y}$ \& $\hat{x}+\hat{z}$ directions respectively. With increasing correlation (U), the SHC for Tb\textsubscript{2}Ir\textsubscript{2}O\textsubscript{7} shows non-linear, complex behaviour. For Bi\textsubscript{2}Ir\textsubscript{2}O\textsubscript{7} we predict a SHC of $-122 \,\, \text{to} \, -512 \, \hbar/2e \, S/cm$.

\begin{figure}
    \centering
    \includegraphics[scale=0.65]{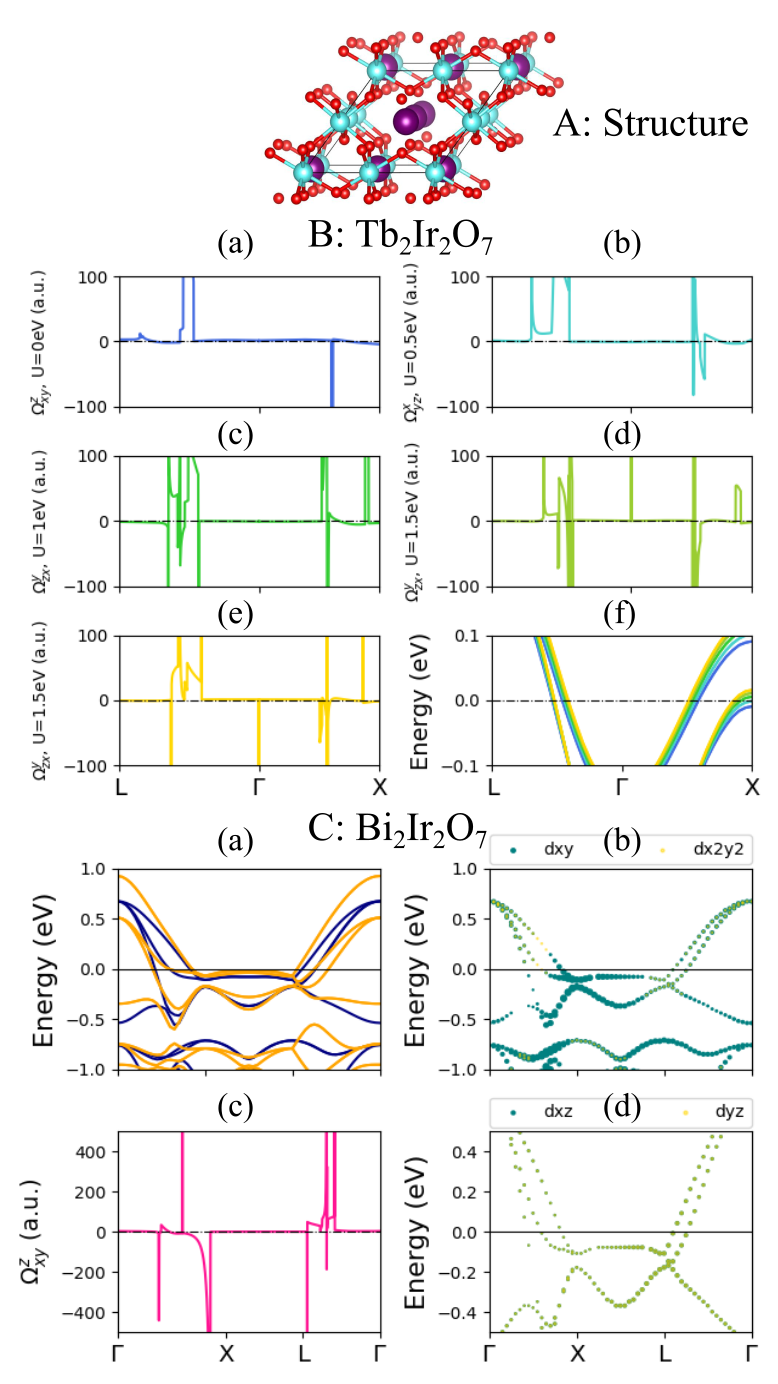}
    \caption{SHE in pyrochlore oxides including Tb\textsubscript{2}Ir\textsubscript{2}O\textsubscript{7} (panel B), Bi\textsubscript{2}Ir\textsubscript{2}O\textsubscript{7} (panel C). Panel A displays the structure with Tb/Bi in cyan, Ir in purple and O in red. For panel B, sections (a)-(e) display the spin Berry curvature plots for Tb\textsubscript{2}Ir\textsubscript{2}O\textsubscript{7} with correlation (U) values ranging from 0 to 1.5 eV and (f) portrays the bandstructure corresponding to the varying U values. For panel C, section (a) shows the bandstructure of Bi\textsubscript{2}Ir\textsubscript{2}O\textsubscript{7} without (blue) and with (orange) SOC, (b) displays the bandstructure projected onto $d_{xy}$ orbitals (green) and $d_{x^2-y^2}$ orbitals (yellow) without SOC, (c) plots the spin Berry curvature and section (d) shows the bandstructure projected onto $d_{xz}$ orbitals (green) and $d_{yz}$ orbitals (yellow) without SOC.}
    \label{fig:5}
\end{figure}

The complicated behaviour of SHC with correlations (U) observed in Tb\textsubscript{2}Ir\textsubscript{2}O\textsubscript{7}, arises due to competing spin Berry curvature hot spots and changes in local band occupation caused by U. At the Fermi level, there exist two degeneracy points in the band structure of Tb\textsubscript{2}Ir\textsubscript{2}O\textsubscript{7} (see Fig. \ref{fig:5}B: f), the first between L and $\Gamma$ and the second between $\Gamma$ and X. Turning on SOC breaks these degeneracies, opening up a gap and generating two spin Berry curvature hot spots (see Fig. \ref{fig:5}B: a-e). These two hot spots give two separate contributions to the overall SHC. For $\sigma^z_{xy}$ and $\sigma^x_{yz}$ the two contributions to each usually oppose one another, with the first, $L-\Gamma$ hot spot contribution being larger than the second, $\Gamma-X$ hot spot contribution. With increase in U, the first contribution generally increases and the second contribution initially increases then decreases (see Fig. \ref{fig:5}B: a-e). Thus, we see  $\sigma^z_{xy}$ and $\sigma^x_{yz}$ first fall with increase of U and then rise. However for $\sigma^y_{zx}$, the two contributions usually add together constructively to yield a negative $\sigma^y_{zx}$, that first rises in value and then falls. The actual picture is more nuanced, but this explanation captures the essential trend. 

The behaviour of the hot spot contributions with U is a result of local band fillings. For the first hot spot ($L-\Gamma$) the band fillings remain essentially unchanged as U increases (see Fig. \ref{fig:5}B: f). Therefore, in general, increasing U increases the SHC contribution from this hot spot. However, for the second hot spot ($\Gamma-X$) the band filling changes substantially for one of the bands as U increases. When U is small the Fermi level goes cleanly through the gap created by SOC, which should maximize the total Berry curvature contribution from this hot spot (see Fig. \ref{fig:5}B: f). When U is increased, the lower band is pushed just above the Fermi level, thereby decreasing the SHC contribution. Therefore, the second hot spot contribution initially increases and then decreases with increasing U. Overall, moderate electron correlations tend to enhance SHE. However, correlations can also suppress SHE in cases where there exist competing spin Berry curvature hot spots or where correlations alter the local band occupation. These inferences are encapsulated in condition 5 (\textit{C5}) of our rational design principles. 

\section{Discussion}
In this paper, we present a comprehensive theory of the foundational factors that determine the value of intrinsic SHE in TMOs. SHE is of fundamental interest owing to its rich physics. It is also highly attractive for enabling energy-efficient, next-generation memory technology. Therefore, our results not only bring important insight into interesting, fundamental physics, they also carry significant promise for helping emerging memory technology. Our findings partly originate from the key realization that intrinsic SHE is inherently a geometric property. Such geometric properties are maximized when the rotation of the degrees of freedom of a wavefunction is maximized. This condition is fulfilled under weak crystal fields and structural distortions. Therefore, we report that weak crystal fields and structural distortions are factors that enhance the value of SHE. Moreover, we derive three other factors that control SHE values, i.e., optimal positioning of the Fermi level, mixing of \textit{J}\textsubscript{eff} = 1/2 \& 3/2 bands, and electron correlations. Our design principles are \textit{general} enough to apply to transition metal compounds, as well as, to other geometric properties like the anomalous Hall effect, orbital Hall effect etc. Our design principles are also impactful because of the inherent tunability of transition metal compounds. In the future, our design principles could be used to develop the highly attractive, \textit{in-situ}, external field control of SHE. 

We also report that new materials (BCC-Pt\textsubscript{3}O\textsubscript{4} and 5d-anti-perovskites) promise to demonstrate \textit{giant} efficiencies of conversion of charge current to spin current, that are an order of magnitude larger than that reported for any oxide so far. Additionally, we enumerate large efficiency values found in a significant number of other oxides. Our discoveries of exciting spin Hall materials along with the implementation of our rational design principles could significantly advance the spin Hall values observed till date. Put together, our work answers fundamental questions about the physics of SHE, it discovers attractive new spin Hall materials, it opens further avenues of materials research for giant SHE, and leads to new possibilities for the development of transformative memory devices with external field control of SHE.

\section{Methods}
We performed calculations using QUANTUM ESPRESSO \cite{Quant_Espresso}, WANNIER90 \cite{Wannier90} and our in-house code which calculates spin Hall conductivity from the output of the former two codes. Previously, we have successfully used this method for the prediction of spin Hall conductivity in lanthanides \cite{Reynolds_lanthanides}. Density functional theory calculations were carried out using QUANTUM ESPRESSO in order to obtain the electronic ground states for our TMOs. Convergence of total energy was ensured for every material, which required a plane-wave energy cut off between 160-200 Ry, and a k-mesh between 5x5x5 and 15x15x15, depending on the structure. Initial structures for these materials were taken from the following sources: BCC-Pt\textsubscript{3}O\textsubscript{4} \cite{Galloni_Pt3O4}, SC-Pt\textsubscript{3}O\textsubscript{4} \cite{Muller_Pt3O4}, Yb\textsubscript{3}PbO \cite{osti_1188128}, BaOsO\textsubscript{3} \cite{Shi_High_pressure}, SrOsO\textsubscript{3} \cite{Shi_High_pressure}, SrIrO\textsubscript{3} \cite{Zhao_ High_pressure}, Tb\textsubscript{2}Ir\textsubscript{2}O\textsubscript{7} \cite{Lefrancois_anisotropy} and Bi\textsubscript{2}Ir\textsubscript{2}O\textsubscript{7} \cite{Qi_strong_Bi2Ir2O7}. Structural optimization was subsequently performed to find the lowest energy structure. 

We utilized norm-conserving, fully relativistic pseudopotentials in the local density approximation (LDA) \cite{Perdew_Accurate, Perdew_Self-interaction}. These pseudopotentials were constructed using the atomic pseudopotential engine \cite{Oliviera_generating} and were benchmarked against the fully relativistic all-electron potential. For an accurate estimation of SHC while using a k-mesh density small enough to be computationally viable, we mapped our DFT ground-state wave functions onto a maximally localized Wannier function basis using WANNIER90. Following this change of basis, we employed an adaptive k-mesh strategy inspired by \cite{Wang_Abinitio_AHE} and employed our in-house code to extract the matrices relevant to the calculation of SHC. To analyze our results, we projected the bandstructures onto relevant atomic orbitals. For bandstructure calculations that included SOC, QUANTUM ESPRESSO did not give us the projection of orbital contributions onto the uncoupled spin orbital states. Therefore, for all materials except Yb\textsubscript{3}PbO, we used Vienna Ab-initio Simulation Package (VASP)\cite{VASP1,VASP2,VASP3} as well as QUANTUM ESPRESSO to analyse the projected bandstructures. The projected bandstructures obtained from VASP were benchmarked with those generated by QUANTUM ESPRESSO for calculations without SOC. Due to difference in the pseudopotential cores for the rare-earth Yb atom, the VASP and QUANTUM ESPRESSO projected bandstructures did not give the same results for Yb\textsubscript{3}PbO. Here, we would like to note that often in the manuscript we have utilized modifiers to describe the strength of spin Hall angle, such as, giant, large etc. These terms are often vague and their meaning changes with every discovery. In this manuscript, we have used the term colossal to describe $\Theta$\textsubscript{SH} $> 20$, giant for $\Theta$\textsubscript{SH} between 2 and 20 and large for $\Theta$\textsubscript{SH} between 0.2 and 2.0.

\section{acknowledgments}
The authors would like to thank Dan Ralph for useful discussions and NSF for financial support under the grants NNCI ECCS-1542159, EFRI-newLAW and NASCENT ERC. They also acknowledge the Texas Advanced Computing Center (TACC) at The University of Texas at Austin for providing HPC resources that have contributed to the research results reported within this paper. URL: http://www.tacc.utexas.edu

\section{Supplementary Materials}

Section I. Justification for neglecting the spin mixing contribution to SHE

Section II. Spin Hall effect in rutile binary oxides

Secion III. Spin Hall effect in other \textit{5d} TMOs

\subsection{Justification for neglecting the spin mixing contribution to SHE}
Our rational design principles for large SHE have been derived under the assumption that in materials with inversion and time-reversal symmetry, the dominant contribution to SHE arises from the spin conserving part of the SOC ($\lambda \hat{l}_z\hat{s}_z$). This dominance of $\lambda \hat{l}_z\hat{s}_z$ has been observed in heavy metals like Pt \cite{Kontani_Pt_ISHE}. Here we give an intuitive argument to show that this assumption is usually, though not always, valid. As shown in the main text, the anomalous velocity originating from the spin conserving part of SOC ($\lambda \hat{l}_z\hat{s}_z$) is $\hat{v}^z_\alpha$ for spin up and $-\hat{v}^z_\alpha$ for spin down. These two terms are always equal in value and opposite in sign, independent of the orbital character of the states. The contribution to spin Berry curvature originating in SOC, in particular the spin conserving part of SOC, is of the form $-\text{Im}[s_{nl}v^z_{\alpha,lm}v^z_{\beta,mn}]$, for say a spin up state. Then the corresponding term for the time reversed, spin down state would be  $-\text{Im}[-(s_{nl})^*-(v^z_{\alpha,lm})^*-(v^z_{\beta,mn})^*]$. Using $\text{Im}(A^*B^*C^*)=-\text{Im}(ABC)$, we can show that these two terms will be equal and always add constructively. 

In contrast, for the spin mixing part of SOC ($\lambda (\hat{l}_+\hat{s}_- + \hat{l}_-\hat{s}_+)$), the anomalous velocity terms for spin up ($\hat{v}^-_\alpha$) and spin down ($\hat{v}^+_\alpha$) are not always exactly equal and opposite. Therefore, depending on the orbital character of the states, the contribution to SHC originating in these terms can interact constructively or destructively. When the contributions from these spin\textit{mixing} terms interact destructively, it is the spin \textit{conserving} part of SOC that gives the dominating term to SHC. Even when the spin \textit{mixing} terms interact constructively, the spin \textit{conserving} contribution can still be important. 

\subsection{Spin Hall effect in rutile binary oxides} 

\begin{table}[h!]
\centering
\begin{tabular}{ ||p{3cm}||p{1.5cm}|p{1.5cm}|p{1.5cm}||p{2cm}||p{1.5cm}|p{1.5cm}|p{1.5cm}||}
 \hline \hline
 \multicolumn{8}{||c||}{Table SI: List of SHC $\sigma^s_{\alpha\beta}$ ($\hbar/2e \, S/cm$), longitudnal conductivity $\sigma$ and spin Hall angle $\Theta^s_{\alpha\beta}$}\\
 \hline
 Structure& $\sigma^z_{xy}$&$\sigma^x_{yz}$&$\sigma^y_{zx}$&$\sigma$ (S/cm)&$\Theta^z_{xy}$& $\Theta^x_{yz}$& $\Theta^y_{zx}$ \\
 \hline \hline
\multicolumn{8}{||c||}{Spin Hall predictions for rutile binary oxides}\\
\hline OsO\textsubscript{2}&     338&	-1152&	-950&	$1.7 \times 10^4$ \cite{Rogers_OsO2}&	0.02&	-0.07&	-0.07\\
\hline IrO\textsubscript{2}&     68&     -288&	-630&	$5.0 \times 10^3$ \cite{Fujiwara_IrO2}&	0.01&	-0.06&	-0.13\\
\hline \hline
\end{tabular}
\end{table}

Similar to the cubic perovskites, rutile binary oxides such as  OsO\textsubscript{2} \cite{Hayakawa_electronic_1999, Mattheiss_electronic_1976} and IrO\textsubscript{2} \cite{Mattheiss_electronic_1976, Panda_effect_2014, Ping_electronic_2015}, also are composed of a transition metal atom inside an octahedral crystal field. This strong octahedral crystal field violates condition \textit{C1}, however it displays a small orthorhombic distortion. To account for correlation effects in IrO\textsubscript{2}, we use an LDA+U scheme with U = 2 eV which is taken from \cite{Panda_effect_2014}. From our analysis below, it is likely that this distortion contributes to SHC via mixing of \textit{J}\textsubscript{eff} = 1/2 \& 3/2 states, satisfying condition \textit{C5}. Together, these conditions result in a spin Hall conductivity of $\sim 10^2 - 10^3 \, \hbar/2e \, S/cm$ for OsO\textsubscript{2} and $\sim 10^2 \, \hbar/2e \, S/cm$ IrO\textsubscript{2}. The values of SHC and $\Theta$\textsubscript{SH} for these materials are calculated along the pseudo-cubic axes and are enumerated in Table SI.

\begin{figure}[h!]
    \centering
    \includegraphics[scale=0.75]{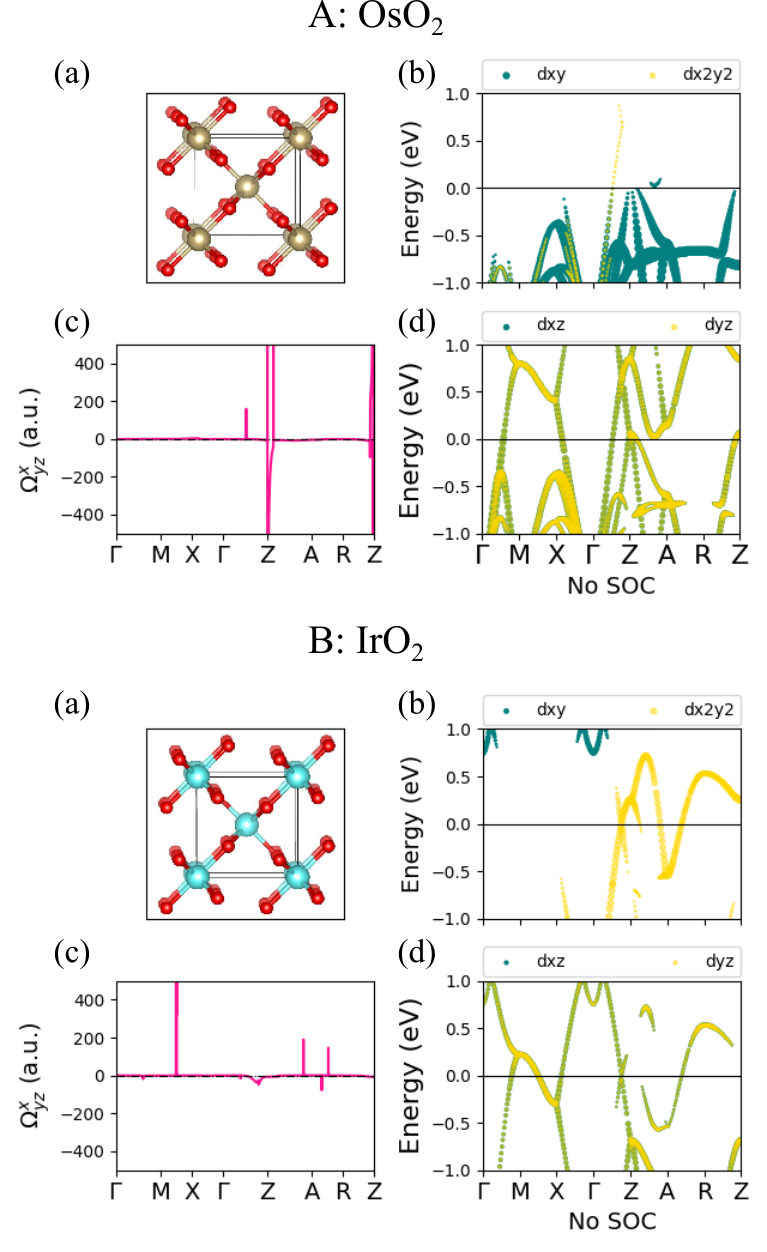}
    \caption{Analysis of SHE in rutile binary oxides, including OsO\textsubscript{2} (panel A) and IrO\textsubscript{2} (panel B). Section (a) displays the rutile structure of these oxides with Os in yellow, Ir in cyan and O in red, (b) shows the bandstructure projected onto $d_{xy}$ orbitals (green) and $d_{x^2-y^2}$ orbitals (yellow), (c) portrays the spin Berry curvature and (d) plots the bandstructure projected onto $d_{xz}$ orbitals (green) and $d_{yz}$ orbitals (yellow).}
    \label{fig:4}
\end{figure}

Acting alone, an octahedral crystal field is expected to split the \textit{5d} bands into a lower energy $t_{2g}$, and a higher energy $e_g$ manifold. The orthorhombic distortion breaks the degeneracy within the $t_{2g}$ manifold by pushing the $d_{xy}$ bands below the degenerate $d_{xz}$ and $d_{yz}$ bands (see Fig. \ref{fig:4}A,B: b,d). The Fermi energy for both materials passes through the lower $t_{2g}$ manifold, with IrO\textsubscript{2} having one more filled electron state per primitive unit cell than OsO\textsubscript{2}. At the Fermi level, there exists a band crossing at the \textit{Z} point in OsO\textsubscript{2} (see Fig. \ref{fig:4}A: b,d) and close to the \textit{Z} point, as well as in the $M-X$ region, in IrO\textsubscript{2} (see Fig. \ref{fig:4}B: b,d). These band crossings give rise to hot spots for spin Berry curvature (see Fig. \ref{fig:4}A,B: c). To analyze the mechanism by which SHE is generated in these rutile oxides, we examine the projected bandstructures around the spin Berry curvature hot spot, as displayed in Fig. \ref{fig:4}A,B: b,d. Around the hot spots, bands for both materials have a strong presence of $d_{xz}$ and $d_{yz}$ orbitals, suggesting that SHE arises primarily from $d_{xz} - d_{yz}$ transitions. The orthorhombic distortions are expected to mix the \textit{J}\textsubscript{eff} = 1/2, 3/2 states \cite{Panda_effect_2014}, which satisfies condition \textit{C4}, and has the effect of enhancing $d_{xz} - d_{yz}$ transitions. A small mixing of these \textit{J}\textsubscript{eff} states has been shown to occur for IrO\textsubscript{2} \cite{Panda_effect_2014}. 

Recently, a study of SHE in IrO\textsubscript{2} \cite{Sun_Dirac_NL} has found Dirac nodal lines (DNLs) \cite{Fang_topological_2015, Chen_topological_2016} to contribute towards SHE. For IrO\textsubscript{2}, a Dirac nodal line (DNL) is shown to occur in the segment of the band that extends in the $M - X$ region at $E_F$. Our findings of a spin Berry curvature hot spot in the $M - X$ region corroborates the results of \cite{Sun_Dirac_NL}. It is interesting to note that DNLs have a direct relationship with SHC, as shown by the authors of \cite{Sun_Dirac_NL}. Dirac nodal lines consist of a line of nodes, where each node is a topological singularity, such that an integral of the Berry curvature over a loop containing a Dirac node is equal to $\pi$ \cite{Fang_topological_2015, Chen_topological_2016}. At points of crossing between bands of differing orbital character, SOC can mix these bands and create spin Berry curvature hot spots that are also Dirac nodes. In such cases, DNLs are essentially a series of spin Berry curvature hot spots that contribute to SHE, as shown for IrO\textsubscript{2} \cite{Sun_Dirac_NL}. An interesting consequence of this relationship is that our rational design principles for large SHE can also be applied to encourage the formation of Dirac nodal lines in TMOs.

\subsection{Spin Hall effect in other \textit{5d} TMOs}
We also list our predictions for SHE in other TMOs in Table SII. Initial structures for the TMOs included in the supplementary were taken from the following sources: OsO\textsubscript{2} \cite{Hayakawa_electronic_1999}, IrO\textsubscript{2} \cite{Bolzan_Structural}, WO\textsubscript{2} \cite{Palmer_Tungsten}, NaWO\textsubscript{3} \cite{osti_1194248}, ReO\textsubscript{3} \cite{Myron_study}, ReBiO\textsubscript{4} \cite{osti_1201578}, Cd\textsubscript{2}Re\textsubscript{2}O\textsubscript{7} \cite{Donohue_Cd2Re2O7}, CdPt\textsubscript{3}O\textsubscript{6} \cite{Prewitt_synthesis}, NaPt\textsubscript{3}O\textsubscript{4} \cite{Waser_NaPt3O4} and PbPt\textsubscript{2}O\textsubscript{4} \cite{Tancret_Synthesis}.

\begin{table}[h!]
\centering
\begin{tabular}{ ||p{3cm}||p{3cm}|p{3cm}|p{3cm}||}
 \hline \hline
 \multicolumn{4}{||c||}{Table SII: List of spin Hall conductivity $\sigma^s_{\alpha\beta}$ ($\hbar/2e \cdot S/cm$)}\\
 \hline
 Structure& $\sigma^z_{xy}$&$\sigma^x_{yz}$&$\sigma^y_{zx}$\\
 \hline
\hline WO\textsubscript{2}&      -2&	    -120&	78\\
\hline NaWO\textsubscript{3}&    -96&	-126&	-126\\
\hline ReO\textsubscript{3}&     -150&	-150&	-128\\
\hline ReBiO\textsubscript{4}&   -168&	-78&	182\\
\hline Cd\textsubscript{2}Re\textsubscript{2}O\textsubscript{7}&   -358&	-374&	-236\\
\hline NaPt\textsubscript{3}O\textsubscript{4}&     204&	360&	300\\
\hline PbPt\textsubscript{2}O\textsubscript{4}&     -12&    284&	-52\\
\hline \hline
\end{tabular}
\label{table:4}
\end{table}

\end{document}